\pdfoutput=1
\documentclass[aps,prl,reprint,showpacs,superscriptaddress]{revtex4-1}
\usepackage[latin1]{inputenc}
\usepackage[T1]{fontenc}
\usepackage[english]{babel}
\usepackage{amsmath}
\usepackage[pdftex]{graphicx}
\usepackage{epstopdf}
\usepackage{SIunits}
\usepackage{multirow}

\usepackage{booktabs}

\begin{document}

\title{Rheology of human blood plasma: Viscoelastic versus Newtonian behavior} 

\author{M. Brust}
\author{C. Schaefer}
\author{R. Doerr}
\affiliation{Experimentalphysik, Universit\"at des Saarlandes, Postfach 151150, 66041 Saarbr\"ucken, Germany}
\author{L. Pan}
\author{M. Garcia}
\author{P. E. Arratia}
\affiliation{Department of Mechanical Engineering \& Applied Mechanics, University of Pennsylvania, Philadelphia, PA 19104, United States}
\author{C. Wagner}
\email{c.wagner@mx.uni-saarland.de}
\homepage{http://agwagner.physik.uni-saarland.de}
\affiliation{Experimentalphysik, Universit\"at des Saarlandes, Postfach 151150, 66041 Saarbr\"ucken, Germany}

\date{\today}

\begin{abstract}
We investigate the rheological characteristics of human blood plasma in shear and elongational flows. While we can confirm a Newtonian behavior in shear flow within experimental resolution, we find a viscoelastic behavior of blood plasma in the pure extensional flow of a capillary break-up rheometer. The influence of the viscoelasticity of blood plasma on capillary blood flow is tested in a microfluidic device with a contraction-expansion geometry. Differential pressure measurements revealed that the plasma has a pronounced flow resistance compared to that of pure water. Supplementary measurements indicate that the viscoelasticity of the plasma might even lead to viscoelastic instabilities under certain conditions. Our findings show that the viscoelastic properties of plasma should not be ignored in future studies on blood flow.
\end{abstract}


\maketitle

Blood is a complex fluid that consists of a suspension of blood cells in a liquid plasma which contains mostly water as well as proteins, mineral ions, hormones, and glucose. In humans, red blood cells (RBC) are the most abundant type of cells in whole blood with a concentration of approximately 45\% by volume. Because of this high RBC concentration, it is often believed that rheological behavior of whole blood is mostly determined by the presence of the RBCs. Blood exhibits shear thinning and at low shear rates the red blood cells form reversible aggregates (rouleaux) that are broken up at high shear rates~\cite{Baskurt2003}. The rouleaux formation is caused by the plasma proteins (most likely due to a depletion effect) but the plasma solution, which is approximately 92\% water, is currently believed to be Newtonian \cite{Baskurt2003,Sousa2011,Wells1961}. For example, experiments by Copley and King \cite{Copley1972} and Jaishankar, Sharma and McKinley \cite{Jaishankar2011} show that the non-Newtonian and viscoelastic effects found in plasma and BSA (Bovine Serum Albumin) solutions in \emph{shear} flows can be attributed to the surface layer of plasma proteins present at the liquid-air interface; that is, these non-Newtonian effects are surface rather than bulk effects. These surface effects can be suppressed by the addition of a small amount of surfactants to the protein solutions~\cite {Jaishankar2011}.

Much work has been devoted to the understanding of laminar blood flow as well as the appearance of flow instabilities that may appear under pathological conditions, e.g.\ near an aneurysm or blockage \cite{Sforza2009,Baek2010}. In general, one has to distinguish between inertia and viscoelastic driven instabilities. An elegant way to separate these two kinds of instabilities is the observation of vortices in a contraction flow device.  Inertial effects lead to downstream vortices whereas upstream vortices can be observed for viscoelastic fluids~\cite{Rodd2005,Rodd2007}.  This approach has been used  in a recent work by Sousa et al.\ \cite{Sousa2011} which gives an overview of the flow of two common blood analog solutions in contraction-expansion microfluidic devices. The two investigated plasma replacement solutions with the same shear but different elongational viscosity revealed a remarkably different flow behavior which shows that it is not sufficient to consider only the shear viscosity in order to build a plasma analog solution. A pronounced viscoelastic behavior in elongational flow has been found in several other bio-fluids, too, e.g.\ for saliva \cite{Bhat2010,Haward2011} or DNA solutions.

\begin{figure}[h!]
\begin{center}
\includegraphics[trim=0 20 0 5,width=0.75\columnwidth]{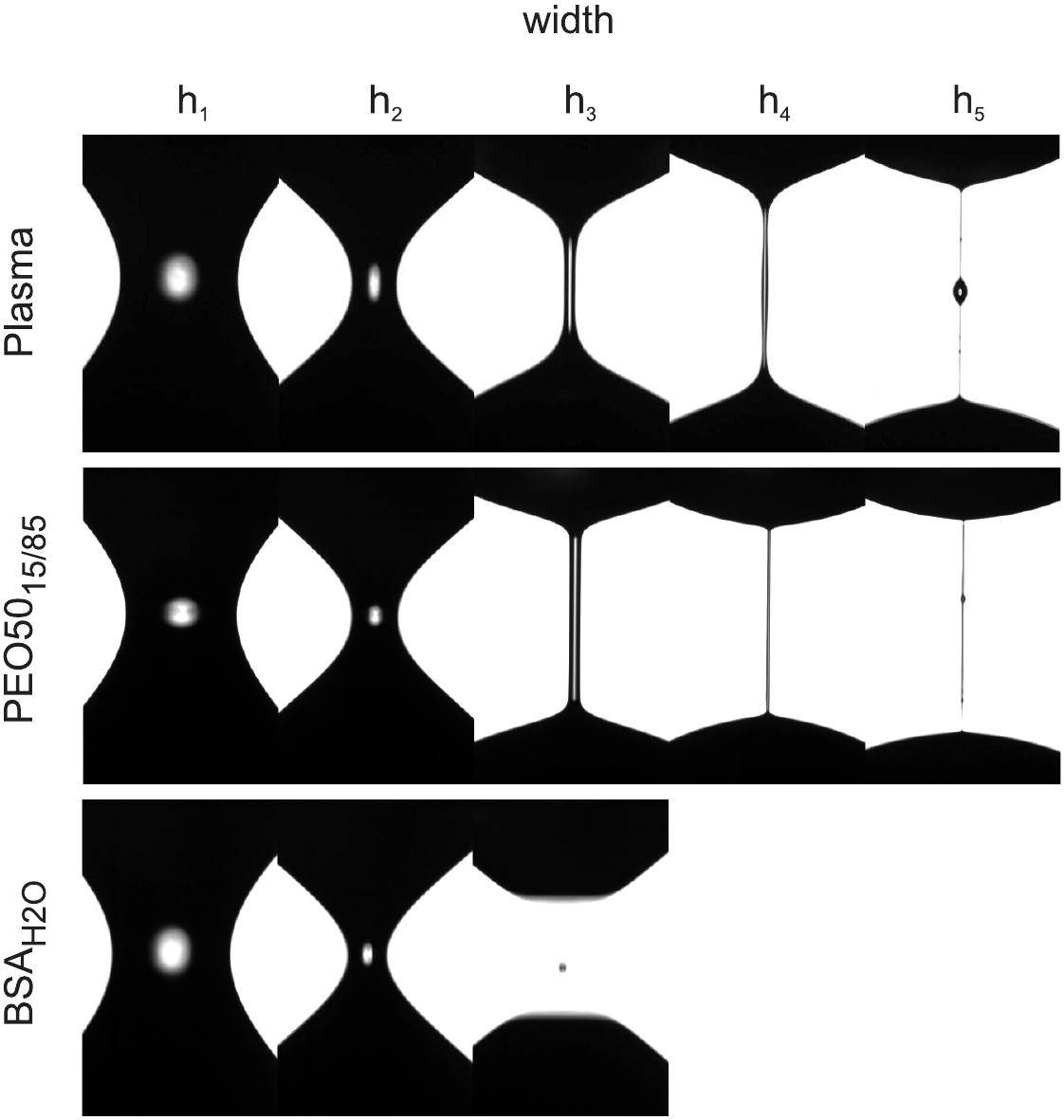}
\end{center}
\caption{\label{fig:thinning} Snapshots of the capillary bridges in the CaBER experiment. Images are $ 833\ \micro\meter \times 1344\ \micro\meter$.}
\end{figure}

\begin{figure}[h!]
\centering
\includegraphics[trim=0mm 10mm 0mm 20mm,width=\columnwidth]{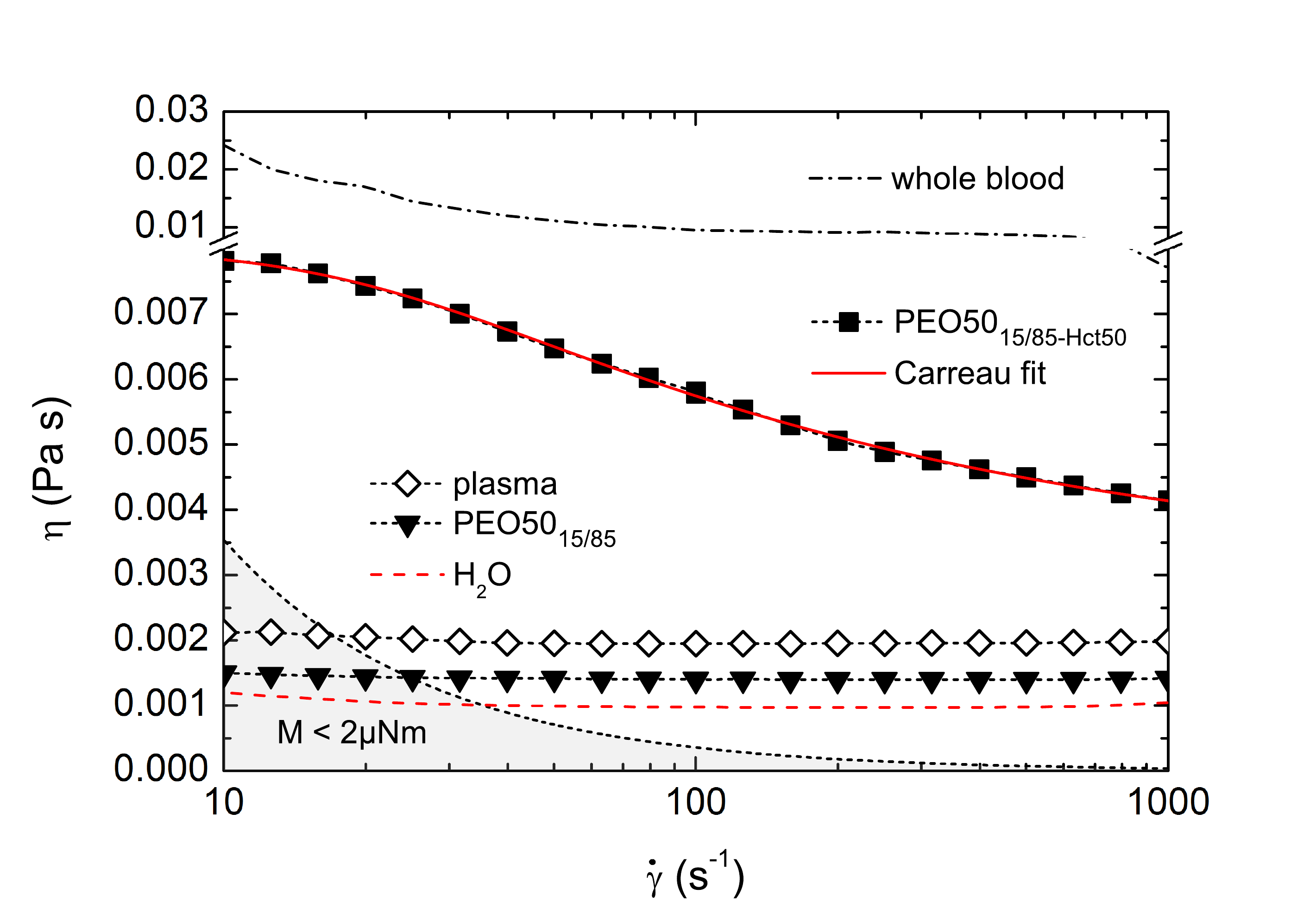}
\caption{(Color online) Shear rate dependent viscosity $\eta(\dot\gamma)$ data at $T=20\celsius$ obtained by the rotational shear measurements. The grey shaded area indicates the regime below the resolution of the rheometer.}
\label{fig:eta_plasma}
\end{figure}

In this manuscript, we investigate the rheology and flow behavior of human blood plasma in both shear and extensional flows. We find that blood plasma shows significant viscoelastic effects in elongational flow in both (i) a capillary breakup extensional rheometer (CaBER, Figure\ref{fig:thinning}) and (ii) a microfluidic contraction-extension device. In addition a synthetic (polymeric) plasma replacement solution is identified that matches the shear and extensional rheology of human plasma.

\emph{Sample solutions}: Blood plasma used in the experiments was is taken from three healthy donors and stored in test tubes using anticoagulants (EDTA or Citra). Plasma is obtained by centrifuging the blood samples at 4000 rotations per minute for 5 minutes immediately after withdrawal. The separated transparent liquid phase is centrifuged again and finally the pure plasma phase is extracted. All measurements are performed on the day of donation. Since the main protein in the plasma is albumin, an aqueous solution of $45\ \milli\gram\per\milli\litre$ albumin (BSA, Polysciences) is prepared for comparison. The plasma replacement solutions are prepared by adding small amounts of polyethylene oxide (PEO, Sigma Aldrich, $M_w=4\cdot 10^6 \ \gram\per\mole$) to a Newtonian solvent. Three PEO solutions are prepared: (i) a 50 ppm PEO in $15/85 wt\%$ glycerol/water solution ($PEO50_{15/85}$), (ii) a 500 ppm PEO in $15/85 wt\%$ glycerol/water solution ($PEO500_{15/85}$), and (iii) a 50 ppm PEO in a $55/45 wt\%$ glycerol/water solution ($PEO50_{55/45}$). Two RBC suspension at $50 vol\%$ are also used, one in physiological buffer (PBS, Phosphate Buffered Saline, Invitrogen) with $15 wt\%$ glycerol ($PBS_{15/85-Hct50}$) and the other under the same condition but with the addition of 50 ppm of PEO ($PEO50_{15/85-Hct50}$). Finally, whole blood and distilled water are used as reference fluids.

Steady shear rheology is performed in a rheometer (Thermo Scientific, HAAKE MARS II) at shear rates ranging from $\dot\gamma=1\ s^{-1}$ to $1000\  s^{-1}$ in rate controlled (CR) mode. A double-cone DC60/1° geometry is used in order to avoid the effect of surface viscoelasticity on the measured shear viscosity $\eta(\dot\gamma)$. To reduce the instrumental noise, a special protocol is applied. For each shear rate, the torque signal is integrated over at least a full revolution of the cone. The minimal resolvable torque is $\tau_\text{min}=2\ \micro\newton\meter$ which limits the reliable minimal shear rate. Above, the accuracy is $\Delta \eta = \pm 0.1\ \milli\pascal\second$. The temperature is kept constant at $T = 20 \pm 1\celsius$.


Figure \ref{fig:eta_plasma} shows the  shear viscosity of blood plasma at room temperature ($RT=20\celsius)$, along with viscosity data of whole blood (average of three donors), two dilute PEO solutions, and a reference curve of pure water. Within the reliable data range (above the grey shaded area), the plasma sample as well as the 50 ppm PEO in $15wt\%$ glycerol-water solution show Newtonian behavior, that is constant shear viscosity $\eta$. By contrast the addition of a physiologic amount of RBCs (Hct50\%) significantly increases the viscosity of the PEO solution. The solution exhibits shear thinning and the flow curve can be fitted to the Carreau model \cite{Tanner1985}. The viscosity of full blood is higher than the viscosity of the  $PEO50_{15/85-Hct50}$ solution. It is known that plasma proteins lead to a depletion induced formation of aggregates of RBCs at low shear rates, and this effect might be less pronounced for the PEO solutions \cite{Baskurt2003}. Table \ref{tab:visc_lambda} summarizes the estimated zero shear viscosities $\eta_0$. We should mention that our determination of $\eta_0$ for whole blood is of limited use only because blood is a yield stress fluid (yield point was found to be approximately 5 mPa).

\begin{table}
\footnotesize
\begin{ruledtabular}
\begin{tabular}{cccc}
\multirow{2}{*}{solution} & temperature & shear viscosity & relaxation time \\
& $T (\celsius)$ & $\eta_0 (\milli\pascal\;\second)$ & $\lambda_C (\milli\second)$\\
\hline
Whole Blood & 37 & 16.9 & $7.8 \pm 0.6$ \\
Plasma & 20 & 1.95 & $2.6 \pm 0.2$\\
Plasma & 37 & 1.34 & $1.5 \pm 0.2$\\
$PEO500_{15/85}$ & 20 & 2.47* & $15.7 \pm 0.2$ \\
$PEO50_{15/85}$ & 20 & 1.40 & $2.6 \pm 0.1$  \\
$PEO50_{55/45}$ & 20 & 7.88 & $7.9 \pm 0.2$  \\
$PEO50_{15/85-Hct50}$& 20 & 8.03* & $1.84 \pm 0.04$ \\
\hline
$BSA_{H_2O}$ & 20 & 1.24 & Newt. \\
$PBS_{15/85-Hct50}$  & 37 &7.6*  & Newt. \\
H$_2$O & 20 & 0.97 & Newt. \\
\end{tabular}
\end{ruledtabular}
\caption{\label{tab:visc_lambda} (Zero) shear viscosities $\eta_0$ and CaBER relaxation times $\lambda_C$ of the sample solutions. Values marked with * are obtained by a fit based upon the Carreau model \cite{Tanner1985}.}
\end{table}

\begin{figure}[h!]
\begin{center}
\includegraphics[trim=0 68 0 50,width=1\columnwidth]{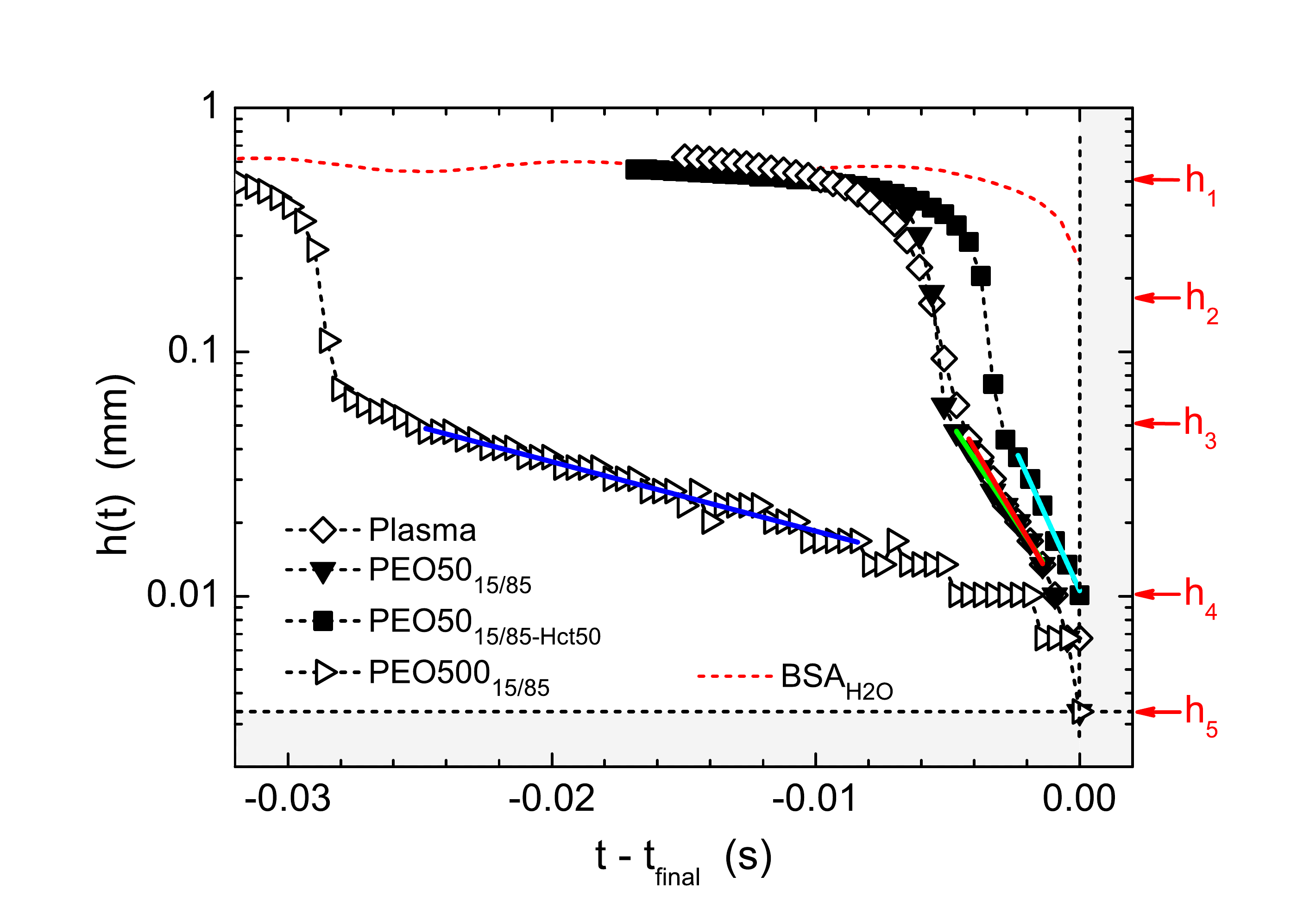}
\end{center}
\caption{\label{fig:CaBER} (Color online) Width $h(t)$ of the capillary bridge as a function of time $t-t_\text{final}$. Markers on the right vertical axis refer to the images shown in Figure \ref{fig:thinning}. Straight lines are exponential fits.}
\end{figure}

The relaxation time of the fluids is obtained using a custom made capillary breakup extensional rheometer (CaBER) setup \cite{Zell2010,Gier2011}. The thinning of the capillary bridge is recorded using a high speed camera and 4x magnification objective  (Pentax, PLAN 4x/0.10, WD=30). We find that the solutions have relatively short relaxation times, and a slow retraction method is used to allow for a quantitative evaluation of the data \cite{Campo-Deano2010}. For a non-Newtonian fluid, the initial inertio-capillar thinning of the capillary bridge is followed by a formation of a straight filament, which thins exponentially in time. The relation $h(t) = h_0 \cdot \exp\left(-(t-t_0)/\lambda_C\right)$ gives the characteristic CaBER relaxation time $\lambda_C$, where $h(t)$ is the diameter of the filament at the time $t$ during the exponential thinning process, which starts at time $t_0$ and width $h_0$ respectively. According to the Oldroyd-B model, a simple approach leads to an estimate for the polymer relaxation time $\lambda=\lambda_C/3$ \cite{Stelter2000,Anna2001}.

Figure \ref{fig:thinning} shows a series of images of the thinning process of the capillary bridges of the plasma as well as  the BSA and $PEO50_{15/85}$ solutions. The capillary bridge of the BSA solution shows the same dynamics as the distilled water (not shown), within experimental resolution. In this case, the capillary bridge breaks fast and leaves a well known satellite droplet. By contrast, the plasma and the $PEO50_{15/85}$ solution form a viscoelastic thread that thins exponentially in time (see Figure \ref{fig:CaBER}). Both solutions also show the typical blistering instability at the very end of the thinning process characterized by the appearance of smaller droplets on the filament \cite{Sattler2008}. These results clearly demonstrate that plasma has viscoelastic properties.

Figure \ref{fig:CaBER} shows the minimal width $h(t)$ of the capillary bridge as a function of time $t-t_\text{final}$. Starting at a width of approximately $h_1\approx0.5\ \milli\meter$, all curves initially describe a uniform thinning down to an intermediate level of $h_2\approx0.17\ \milli\meter$. At this stage, polymers in the solution do not yet affect the flow and the difference between the solutions can be only seen during further thinning of the filament. While the capillary bridge of the Newtonian sample breaks very rapidly (cp.\ width $h_3$), the other samples form a filament which thins exponentially over characteristic time scales $\lambda_C$. Table \ref{tab:visc_lambda} summarizes the corresponding relaxation times as averages over all performed measurements. Note that the BSA solution does not show any non-Newtonian behavior and therefore we can exclude that the protein surface layer of the plasma is responsible for the elastic thinning behavior (in agreement with \cite{{Regev2010}}), but some other plasma proteins in the bulk that have not yet been identified are responsible for the viscoelasticity. This result is also corroborated in a cone-plate geometry with a free surface layer: adding $0.01 wt\%$ of the soluble nonionic surfactant (Tween 20, critical micellar concentration $cmc = 0.07 wt\%$ to the plasma did indeed suppress the apparent shear thinning (data not shown) in agreement with the experiments described in Ref.\ ~\cite {Jaishankar2011}). Please note that in principle surfactants can lead to viscoelastic filaments themselves but in our case the surfactant dissolved in pure water did not show any filament \cite{Craster2005,Craster2009,Liao2006,Liao2004,Kwak2001,McGough2006,Matar2002,Jin2006,Hansen2000}.  However, the addition of a surfactant to the plasma did \emph{not} alter the thinning process in the CaBER which again shows that the filament is not caused by a protein surface layer. We note that additional thinning experiments of the plasma solution were performed in which the surrounding air was replaced by silicon oil and a pronounced filament was as well observed (data not shown).

Again, we can compare our polymer solutions that should serve as model systems with the plasma samples. The data presented in Table \ref{tab:visc_lambda} shows that the $PEO50_{15/85}$ solution is a good plasma replacement solution because it matches the plasma shear and elongational properties; by simply increasing the solvent viscosity of this solution we get the $PEO50_{55/45}$ sample that matches the elongational properties of full blood but it does not match its shear rheology very well. Therefore, we compared a PBS buffer solution with $50vol\%$ hematocrit without polymer ($PBS_{15/85-Hct50}$), with polymer ($PEO50_{15/85-Hct50}$) and full blood. The polymeric hematocrit solutions reflects the elongational properties of blood to some extent. However, the $PBS_{15/85-Hct50}$ solution simply breaks-up like a Newtonian liquid. This clearly shows that the non-Newtonian elongational viscosity (i.e. the elastic properties of the macromolecules) of the plasma contributes to the non-Newtonian elongational viscosity of blood. Again, an additional clustering of the RBC's in the whole blood due to the rouleaux formation might explain the remaining discrepancies.

Next, the effects of this apparent viscoelasticity of plasma flow are investigated in a microfluidic contraction-expansion device by measuring the pressure drop across the contraction part of the geometry \cite{Rodd2005,Rodd2007}. This geometry has a significant elongation flow component and it resembles flow restricted vessels or the flow at a branching from a larger to a smaller vessel. The microchannels are made of PDMS using standard soft lithography methods. Their length, width and depth are $L=30\ \milli\meter$, $W=400\ \micro\meter$ and $h=50\ \micro\meter$ respectively. These rectangular microchannels exhibit sharp contraction of reduced width $w=25\ \micro\meter$ and length $l=100\ \micro\meter$ in the middle. Note that the ratio of the expansion width to the contraction width is 1:16 which sets the (Hencky) elongation strain $\epsilon=\ln(16)\approx2.8$.  Pressure measurements are performed by directly incorporating two pressure taps (Honeywell 26PC) into the microchannel, both upstream and downstream of the contraction plane. The pressure sensors are used to cover a differential pressure range of $0 < \triangle P < 56.5$ kPa. This type of measurements have been successfully applied to investigate the flow of viscoelastic fluids in expansion-contraction microfluidic devices \cite{Rodd2005}.

Results for the pressure drop of both human plasma and water as a function of flow rate, or equivalently Reynolds number ($Re$) are shown in Figure 4. Note that here $Re = \rho V D_h / \eta_0$ where $V$ is the mean flow velocity, $\rho$ the fluid density, $\eta_0$ the zero shear rate viscosity and $D_h=2wh/(w+h)$ the hydraulic diameter.  For example, Figure 4 (inset) shows that the water pressure drop increases linearly as the flow rate is increased. On the other hand, the blood plasma pressure drop shows a deviation from this linear trend. This nonlinear behavior can also be observed by normalizing the pressure drop at a given flow rate by the pressure drop of water at the same flow rate. Figure 4 shows that the dimensionless pressure drop of plasma decreases significantly as the $Re$ increases. This indicates an extensional strain-rate thinning viscosity behavior that has been observed in other types of complex fluids such as micellar solutions \cite{Bhardwaj2007}. These measurements clearly show that blood plasma is a non-Newtonian fluid. We also checked by flow visualization that the increase in resistance is not caused by a viscoelastic instability, but we found that in principle the viscoelastic behavior of plasma could be responsible for viscoelastic instabilities, especially in complex flow situations and in the presence of red blood cells, i.e. $50\%$ hematocrit \cite{SM}.

\begin{figure}[h!]
\centering
\includegraphics[trim=0mm 0mm 0mm 10mm,width=0.95\columnwidth]{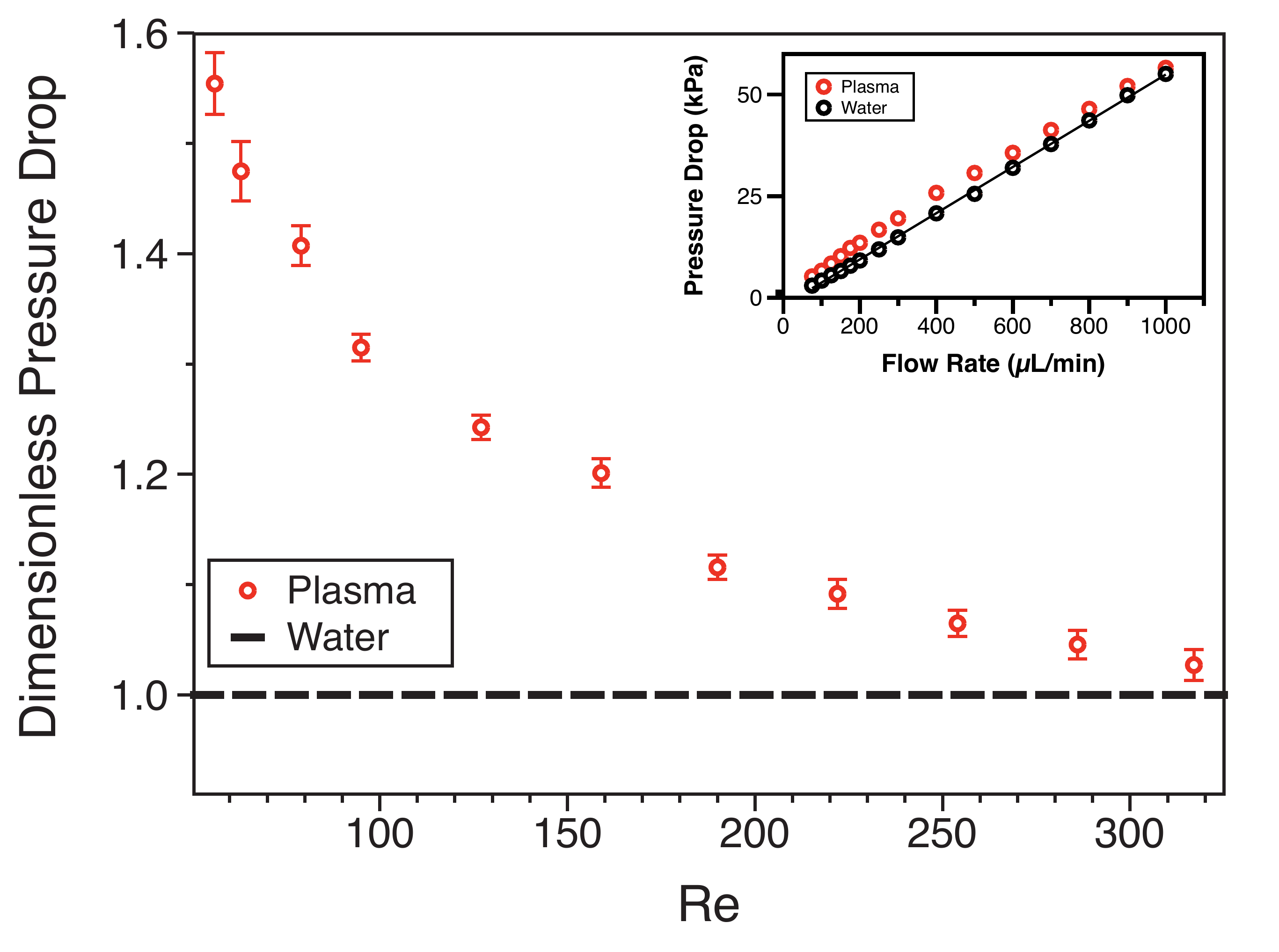}
\caption{(Color online) Dimensionless pressure drop as a function of Reynolds number. The data shows higher values of pressure drop for human plasma compared to water. Blood plasma shows a nonlinear, 'thinning' behavior of pressure drop as a function of strain-rate \cite{Bhardwaj2007}. Inset: Pressure drop as a function of flow rate. Blood plasma shows nonlinear behavior.}
\label{fig:pressure}
\end{figure}

\emph{In conclusion}, we found a pronounced viscoelastic behavior of human blood plasma in a capillary breakup extensional rheometer (CaBER).  Similar to dilute polymer solution, steady shear flow or small amplitude oscillatory shear does not indicate any elasticity of the plasma because it stretches polymers less efficiently. There are several reports on viscoelasticity of plasma in shear flow but they all could be attributed to the formation of a protein surface layer which is not the case in our measurements. We could also show that the elongational properties of blood are to a large extend determined by elongational properties of the plasma proteins.
In order to check the generality of our findings we prepared a plasma replacement fluid with similar elongational and shear properties. Differential pressure measurements in a microfluidic contraction-expansion  geometry confirmed the non-Newtonian character of plasma. Supplementary measurements with flow visualization indicate that this viscoelasticity might even lead to viscoelastic flow instabilities. Similarly, it is known that small amounts of polymers with elongational properties as our plasma replacement solution can lead to turbulent drag reduction, i.e. turbulent instabilities due to high inertia might be suppressed by the elongational properties of the plasma. Finally, recent numerical results indicate that a slight viscoelasticity of the solvent might lead to a pronounced cell depletion layer close to the vessel boundaries \cite{Pranay2011}. In view of these findings, it is expected that the viscoelasticity of plasma has to be taken into account in future studies.

This work was supported by the DFG Graduate college GRK1276 "Stucture formation and transport in complex systems". L. Pan and P. E. Arratia are supported by NSF-CBET-0932449. We thank Anke Lindner and Alexander Morozov for carefully reading the manuscript.


%

\end {document}